# New statistic for financial return distributions: power-law or exponential?

V.Pisarenko[1] and D.Sornette[2,3]

[1]International Institute of Earthquake Prediction Theory and Mathematical Geophysics, Russian Ac. Sci. Warshavskoye sh., 79, kor. 2, Moscow 113556, Russia Federation. E-mail: vlad@sirius.mitp.ru

[2]Institute of Geophysics and Planetary Physics and Department of Earth and Space Science University of California, Los Angeles, California 90095. E-mail: sornette@moho.ess.ucla.edu

[3]Laboratoire de Physique de la Matière Condensée CNRS UMR6622 and Université des Sciences, B.P. 70, Parc Valrose 06108 Nice Cedex 2, France

***Abstract.***

We introduce a new statistical tool (the *TP*-statistic and *TE*-statistic) designed specifically to compare the behavior of the sample tail of distributions with power-law and exponential tails as a function of the lower threshold $u$. One important property of these statistics is that they converge to zero for power laws or for exponentials correspondingly, regardless of the value of the exponent or of the form parameter. This is particularly useful for testing the structure of a distribution (power law or not, exponential or not) independently of the possibility of quantifying the values of the parameters. We apply these statistics to the distribution of returns of one century of daily data for the Dow Jones Industrial Average and over one year of 5-minutes data of the Nasdaq Composite index. Our analysis confirms previous works showing the tendency for the tails to resemble more and more a power law for the highest quantiles but we can detect clear deviations that suggest that the structure of the tails of the distributions of returns is more complex than usually assumed; it is clearly more complex that just a power law.

## 1-Introduction

The distribution of returns is one of the most basic characteristics of the stock markets and many papers have been devoted to it. One could summarize the present situation by saying that a large number of papers document tails of the probability distribution function (PDF) of returns which are heavier than a Gaussian tail and heavier than an exponential tail and are rather well approximated by a power law with exponent in the range 2.5-3.5, such that the existence of the third (skewness) and the fourth (kurtosis) moments is questionable. The fat-tail nature of the PDF is best observed for intra-day time scales (minutes) [1-4]. A few works including two recent papers [5,6] of the present authors (with Y. Malevergne) propose an alternative parameterization in terms of stretched exponentials, based on statistical tests suggesting that the exponent of the power law tail does not seem stable in the tail of the PDFs of datasets of medium size (20000 values) which are typical of standard financial applications. We refer to [1-4] for evidence of power law tails in high frequency data and [5,6] for a review and discussion of alternative models of the PDFs of financial returns.

One important origin behind the controversy on the nature of the tail of PDFs of financial returns stems from the use of different parts $(u,\infty)$ of the tails $1-F(x)$ ($F(x)$ being the cumulative distribution function). Usually, in the lower part of the quantile range, one observes a slow decreasing tail whereas, at the extreme part of the quantile range, the decay rate of the tail becomes considerably faster. Thus, the answer to the question "what is the tail of the observed distribution?" depends to a large extent on the lower threshold $u$ of the data used in fitting procedures. For instance, works documenting the relevance of stable Levy laws use the full sample range [7-9] while discrepancies appear when constraining the analyses to different tail ranges. Malevergne et al. [5,6] show in particular that the apparent power law exponent of the Pareto increases with the quantiles and its growth does not seem exhausted for the highest quantiles of three out of the four tail distributions investigated. The Boston group [1,2] circumvents this difficulty by combining high-frequency data for thousands of companies by normalizing each data set by its standard deviation, thus introducing in this way an effect not completely understood. Or they fit individually the high-frequency PDFs of thousands of companies by a power law and find that most exponents lie in the range 2-4. It is clear that such parametric fits can only address the question of what is the average representation of the tail of PDF (in some range pre-defined rather arbitrarily) by a power law but do not test in details the adequacy of this model, especially as a function of the depth (quantile) in the tail of the PDFs.

Here, we suggest a new statistical tool (that we call *T*-statistic) designed specifically to describe the behavior of the sample tail as compared with power-like and exponential tails as a function of the lower threshold $u$. Specifically, we propose two statistics: $TP(u)$ and $TE(u)$. The former (respectively latter) is asymptotically small (i.e. converges to zero as the sample size $n$ tends to infinity) ***for the power tail described by the Pareto distribution:*** $1-F(x) = (u/x)^b$ ; $x \geq u$ ***with arbitrary power index b*** (respectively ***for the exponential tail*** $1-F(x)=\exp(-(x-u)/d)$; $x \geq u$ ***with an arbitrary form parameter d***). At the same time, $TP(u)$ (respectively $TE(u)$) deviates from zero for distributions differing from the target one, i.e. the Pareto distribution (respectively the exponential distribution). Thus, we can use the *TP*-statistic in order to expose in a very clear, visual and understandable way the deviations of a given sample from the Pareto distribution *with arbitrary power index b* (respectively the exponential distribution with arbitrary form parameter *d*). Of course, the statistical scatter of the *TP* and *TE* statistics should and will be taken into account.

Thus, instead of answering the question "what does the sample tail look like: power-like or exponential-like?" we answer with the help of *TP*- and *TE*-statistics to a rather different question: "***what part of the extreme tail*** looks power-like (or, exponential-like)?" Of course, negative answers to both questions are possible.

We illustrate our introduced *TP*- and *TE*-statistics on two datasets previously studied in [5,6]: the Dow Jones Industrial Average (DJIA) daily log-returns over the twentieth century, and the Nasdaq Composite index (ND) 5-minutes log-returns over one year from April 1997 to May 1998 obtained from Bloomberg (for positive and negative returns separately). Our *T*-statistics shows that all 4 samples tails as functions of the lower threshold *u* behave similarly: in the bulk of the samples, they are close to an exponential distribution while, at the extreme range, they approach a power-like decay. *TP*- and *TE*-statistics provide the possibility of locating (of course with some statistical uncertainty) the value of the "cross-over" or "change-of-regime" between these two limiting behaviors for each corresponding sample. This shows that the risk measures such as VaR (value-at-risk) and expected shortfall cannot be estimated in a standard way with the same distributional model for all quantile or confidence levels. This has important consequences for risk assessment and management. Our *TP*- and *TE*-statistics also show that, while the power law model becomes better than the exponential model in the tails, there exists detectable deviations as a function of the lower thresholds, even for the largest quantiles. Our analysis seems to confirm that the power law tail is not fully sufficient to model in full the asymptotic tails of the distributions of financial returns in datasets of about 20000 data points, whose sizes are significant compared with those used routinely in investment and portfolio analysis.

Section 2 introduces the *TP*-statistics and derives its properties. Section 3 introduces the *TE*-statistics and derives its properties. Section 4 presents the application of the *TP*- and *TE*-statistics to the DJIA and ND data sets. Section 5 concludes.

## 2- *TP*-statistic and its properties

Consider the Pareto distribution $F(x)$, conditioned on the semi-axis $x \geq u$ :

$$F(x) = 1 - (u/x)^b \ , \ x \geq u, \ b > 0, \qquad (1)$$

where $u$ is a lower threshold, and $b$ is the power index of the distribution. Let us consider a finite sample $x_1,\ldots, x_n$. In [15], the following statistic $TP = TP(x_1,\ldots, x_n)$ was suggested such that, asymptotically for large $n$, $TP$ would be close to zero and, at the same time, would deviate from zero for samples whose distribution deviates from eq.(1). This statistic is based on the first two normalized statistical log-moments of the distribution (1). Using the symbol E for the mathematical expectation, we have

$$E_1 \equiv E\ log(X/u) = \int_u^\infty log(x/u)\ dF(x) = 1/b\ ; \qquad (2)$$

$$E_2 \equiv E\ log^2(X/u) = \int_u^\infty log^2(x/u)\ dF(x) = 2/b^2. \qquad (3)$$

Thus, if we choose

$$TP = (1/n) \sum_{k=1}^{n} log(x_k/u) )^2 - (0.5/n) \sum_{k=1}^{n} log^2(x_k/u),$$

then according to the Law of Large Numbers and equations (2)-(3), the statistic *TP* tends to zero as $n \to \infty$ . In order to evaluate the standard deviation *std(TP)* of the statistic *TP*, we rewrite the expression for *TP* in the form:

$$(4) \qquad TP = (1/n) \sum_{k=1}^{n} [log(x_k/u) - E_1] + E_1)^2 - (0.5/n) \sum_{k=1}^{n} [log^2(x_k/u) - E_2] - 0.5E_2,$$

where $E_1$, $E_2$ are the expectations of $log(x_k/u)$ and $log^2(x_k/u)$ respectively (for Pareto samples, $E_1, = 1/b$ and $E_2 = 2/b^2$ as given in expressions (2) and (3)). Both sums in eq.(4) are of the order $n^{-0.5}$:

$$\varepsilon_1 = 1/n \sum_{k=1}^{n} [log(x_k/u) - E_1] \propto n^{-0.5}; \ \varepsilon_2 = 1/n \sum_{k=1}^{n} [log^2(x_k/u) - 2/b^2] \propto n^{-0.5}$$

Thus, if *n* is large enough, we can expand *TP* in eq.(4) into Taylor series up to terms of the order $n^{-0.5}$ in the neighborhood of $E_1$ and $E_2$ respectively:

$$(5) \qquad TP \cong (E^2_1 - 0.5E_2) + 2E_1\varepsilon_1 - 0.5\, \varepsilon_2 .$$

This provides an estimation of *std(TP)* by the standard deviation of the sum:

$$(6) \qquad 2E_1\varepsilon_1 - 0.5\, \varepsilon_2 = (2\, E_1 /n) \sum_{k=1}^{n} [log(x_k/u) - E_1] - (0.5/n) \sum_{k=1}^{n} [log^2(x_k/u) - E_2] =$$

$$= (0.5\, E_2 - 2\, E^2_1) + (1/n) \sum_{k=1}^{n} [2\, E_1\, log(x_k/u) - 0.5\, log^2(x_k/u)].$$

The standard deviation of the last sum in (6) can be estimated by

$$(7) \qquad n^{-0.5}\, std[2\, E_1\, log(x_k/u) - 0.5\, log^2(x_k/u)],$$

and the standard deviation *std* of the term in the bracket in eq.(7) is estimated through its sampled value $[2\, E_1\, log(x_k/u) - 0.5\, log^2(x_k/u)]$. Equation (6) then provides an estimate of *std(TP)* if we replace $E_1$ by its sample analog:

$$(1/n) \sum_{k=1}^{n} log(x_k/u).$$

## 3- *TE*-statistic and its properties

Let us consider the exponential distribution, conditioned to the semi-axis $x \geq u$ :

$$(8) \qquad F(x) = 1 - \exp( -(x-u)/d ), \quad x \geq u, \quad d > 0,$$

where $u$ is a lower threshold, and $d$ is the form parameter (scale parameter) of the distribution. Let us consider a finite sample $x_1,\ldots, x_n$. It is desirable to construct a statistic $TE = TE(x_1,\ldots, x_n)$ such that, asymptotically for large $n$, $TE$ would be close to zero and, at the same time, would deviate from zero for samples whose distribution deviates from eq.(8). Let us construct such statistic based on the first two normalized statistical (shifted) log-moments of the distribution (8). Using again the symbol $E$ for the mathematical expectation, we have

$$(9) \qquad E \log(X/u -1) = \int_u^\infty \log(x/u -1)\, dF(x) = \log(d/u) -C ,$$

where $C$ is the Euler constant: $C = 0.577215\ldots$ .

$$(10) \qquad E \log^2(X/u -1) = \int_u^\infty \log^2(x/u -1)\, dF(x) = ( \log(d/u) -C )^2 +\pi^2/6.$$

Thus, if we choose

$$(11) \qquad TE = 1/n \sum_{k=1}^{n} \log^2(x_k/u -1) - (1/n \sum_{k=1}^{n} \log(x_k/u -1) )^2 - \pi^2/6,$$

then according to the Law of Large Numbers and equations (9)-(10), the statistic $TE$ tends to zero as $n \to\infty$. In order to evaluate the standard deviation $std(TE)$ of the statistic $TE$, we rewrite (11) in the form:

$$(12) \qquad TE = \pi^2/6 + 1/n\sum_{k=1}^{n} [ ( \log(x_k/u-1) -E \log(X/u -1)]^2 .$$

This provides an estimation of $std(TE)$ by the standard deviation of the sum:

$$(13) \qquad 1/n\sum_{k=1}^{n} [ ( \log(x_k/u-1) -E \log(X/u -1)]^2 ,$$

which can be easily performed.

## 4- Application of the *TP*- and *TE*-statistics to the DJIA and ND data sets

This section presents a series of figures plotting the *TP*-statistic *TP(u)* and the *TE*-statistic *TE(u)* as a function of the lower threshold *u* for returns of the DJIA at the daily time-scale and for the ND at the 5-minute time-scale.

In order to appreciate the information contained in these plots, it is useful to test the corresponding statistics for a pure Pareto sample and for a pure exponential sample. Fig.1a shows the *TP*-statistic *TP(u)* as a function of the lower threshold *u*, applied to a simulated Pareto sample of size $n=20000$ with power index $b$ = 3 generated with $u$=1 as defined in Eq.(1). Fig.1b shows the *TE*-statistic *TE(u)* for this pure power law. Fig.2a shows the *TP*-statistic *TP(u)* applied to a simulated exponential sample of size $n=20000$ with form parameter $d$ = 4 generated with $u$=1 as defined in Eq.(8). Fig.2b shows the *TE*-statistic *TE(u)* for this pure exponential law.

Fig.3 shows the *TP*-statistic *TP(u)* and the *TE*-statistic *TE(u)* as a function of the lower threshold *u* for positive returns of the DJIA at the daily time-scale. The *TP(u)* statistic shows clearly a tendency for a convergence towards a power law behavior, but statistically significant departures still remain up to the largest available thresholds. The *TE(u)* statistic can not exclude the exponential model for lower thresholds above 3% but its standard deviations are very large and this statistic thus lacks power of discrimination.

Fig.4 shows the *TP*-statistic *TP(u)* (a) and the *TE*-statistic *TE(u)* as a function of the lower threshold *u* for negative returns of the DJIA at the daily time-scale. The *TP(u)* statistic shows a convergence towards a power law behavior which is accepted only for the largest available threshold $u$>4%. The *TE(u)* statistic can not exclude the exponential model for lower thresholds above 3% but its standard deviations are very large and this statistic thus lacks power of discrimination.

Fig.5 shows the *TP*-statistic *TP(u)* (a) and the *TE*-statistic *TE(u)* as a function of the lower threshold *u* for positive returns of the ND at the 5-minutes time-scale. The *TP(u)* statistic shows a convergence towards a power law behavior over an intermediate range $1\% < u < 3\%$ but shows a distinct departure for a power law further deep in the tail for $u>3\%$. The *TE(u)* statistic clearly excludes the exponential model for all lower thresholds and shows a non-monotonous behavior of the tail, confirming the observation of the *TP(u)* statistic.

Fig.6 shows the *TP*-statistic *TP(u)* (a) and the *TE*-statistic *TE(u)* as a function of the lower threshold *u* for negative returns of the ND at the 5-minutes time-scale. The *TP(u)* statistic shows a clear convergence towards a power law behavior which becomes approximately stable for $0.15\% < u$. A slight departure from the pure power law behavior can be observed for the largest thresholds. The *TE(u)* statistic can not exclude the exponential model for almost all lower thresholds but its standard deviations are very large and this statistic thus lacks power of discrimination.

## 5- Concluding remarks

These analyses complement those presented in [5,6]: they confirm the tendency for the tails to resemble more and more a power law for the highest quantiles but we can detect clear deviations that suggest that the structure of the tails of the distributions of returns is more complex than usually assumed; it is clearly more complex that just a power law. The nature of the real tail of the PDF of financial returns has not been settled here: our conservative conclusion is that, while the power law may present an approximate model, it may be inaccurate to extrapolate such a model beyond the range of observed returns. Being inaccurate, it may be inappropriate to use for risk assessment purposes because the unobserved tail may be thinner or fatter than estimated in the finite available range. Extrapolations are dangerous in view of the distinct deviations from the reference power law or exponential models. We believe in particular that the existence of conditional dependencies for the largest returns (see [10-14] and references therein) may impact significantly on the tail structure for the highest quantiles which may thus take a different asymptotic form than determined in the empirical range. Such intermittent dependences may perhaps be in part at the origin of the complexity of the distribution of financial returns. Such a conclusion has been clearly shown to describe the distributions of drawdowns, i.e., runs of returns from top to bottom prices, which are characterized by two regimes [12,14]: an approximate exponential body and tails much more heavy-tailed that the extrapolation of the exponential law (called "outliers" or "kings").

**FIGURES**

Fig. 1a

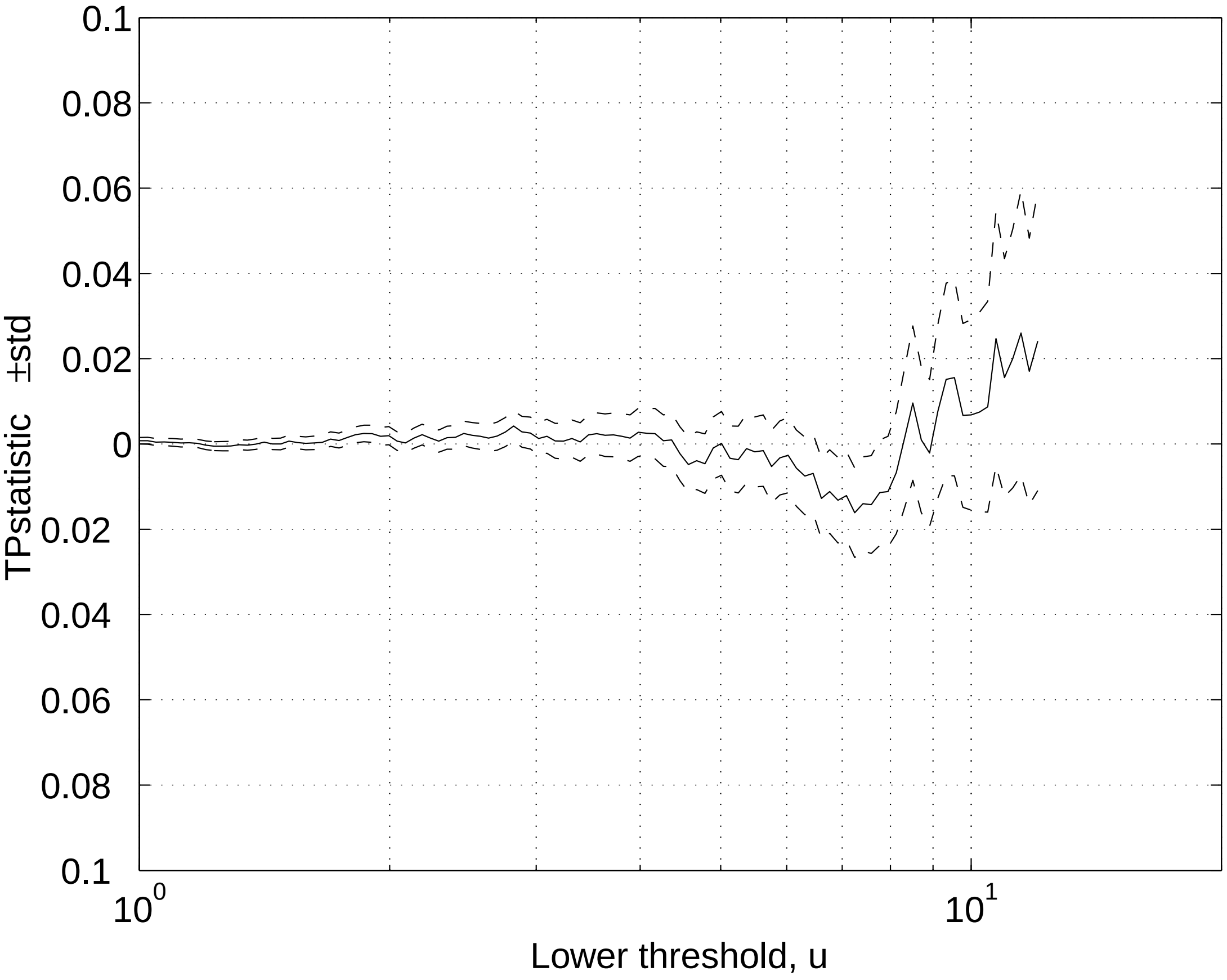

Fig. 1b

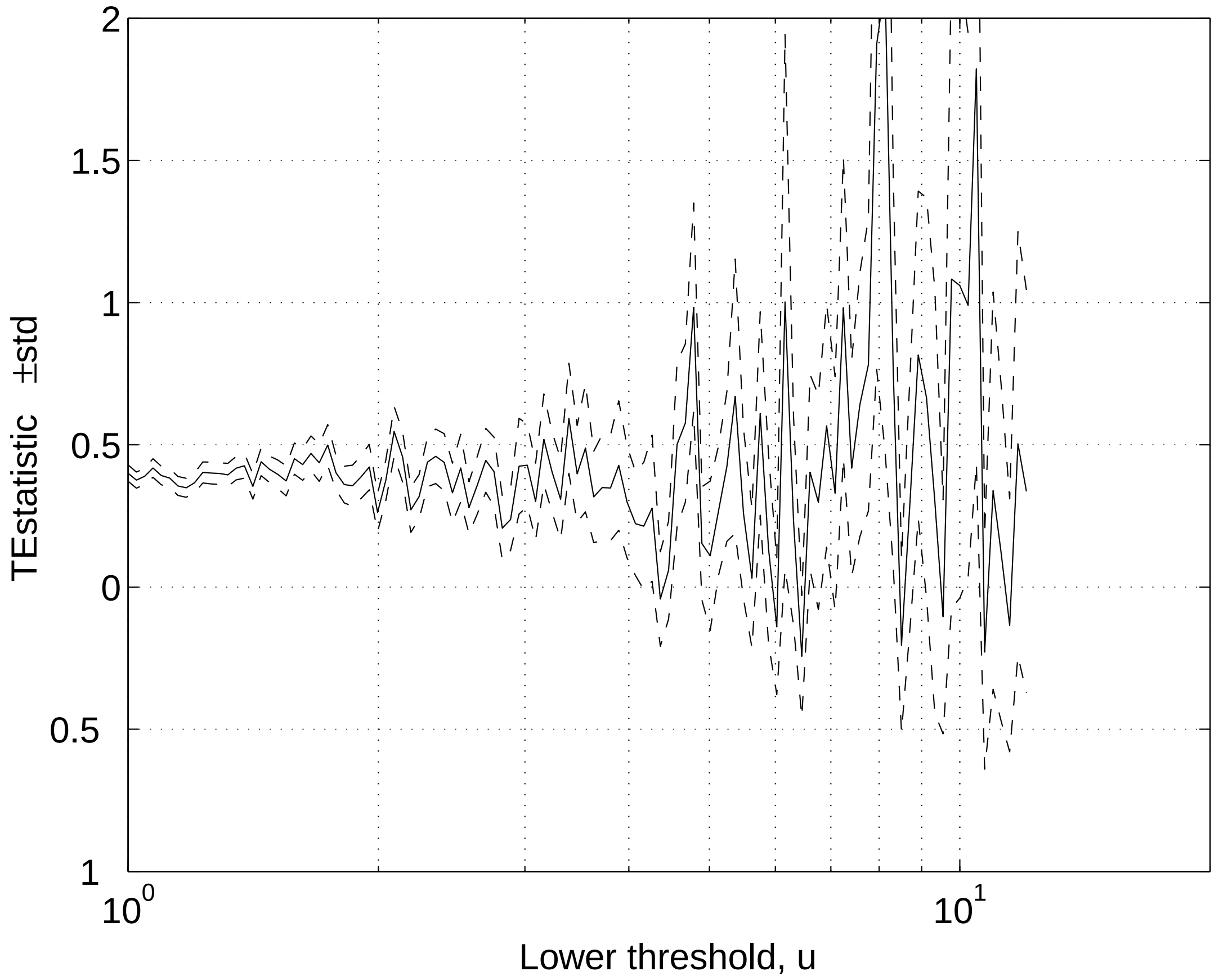

Fig. 1: a) *TP*-statistic as a function of the lower threshold *u*, applied to a simulated Pareto sample of size *n=20000* with power index *b* = 3 generated with *u*=1 as defined in Eq.(1). Increasing *u* decreases the number of data values used in the calculation of the statistic *TP*, thus enhancing the fluctuations around 0. The two dashed lines show plus or minus one standard deviation std estimated as exposed in the text. b) *TE*-statistic of the same synthetic data.

Fig. 2a

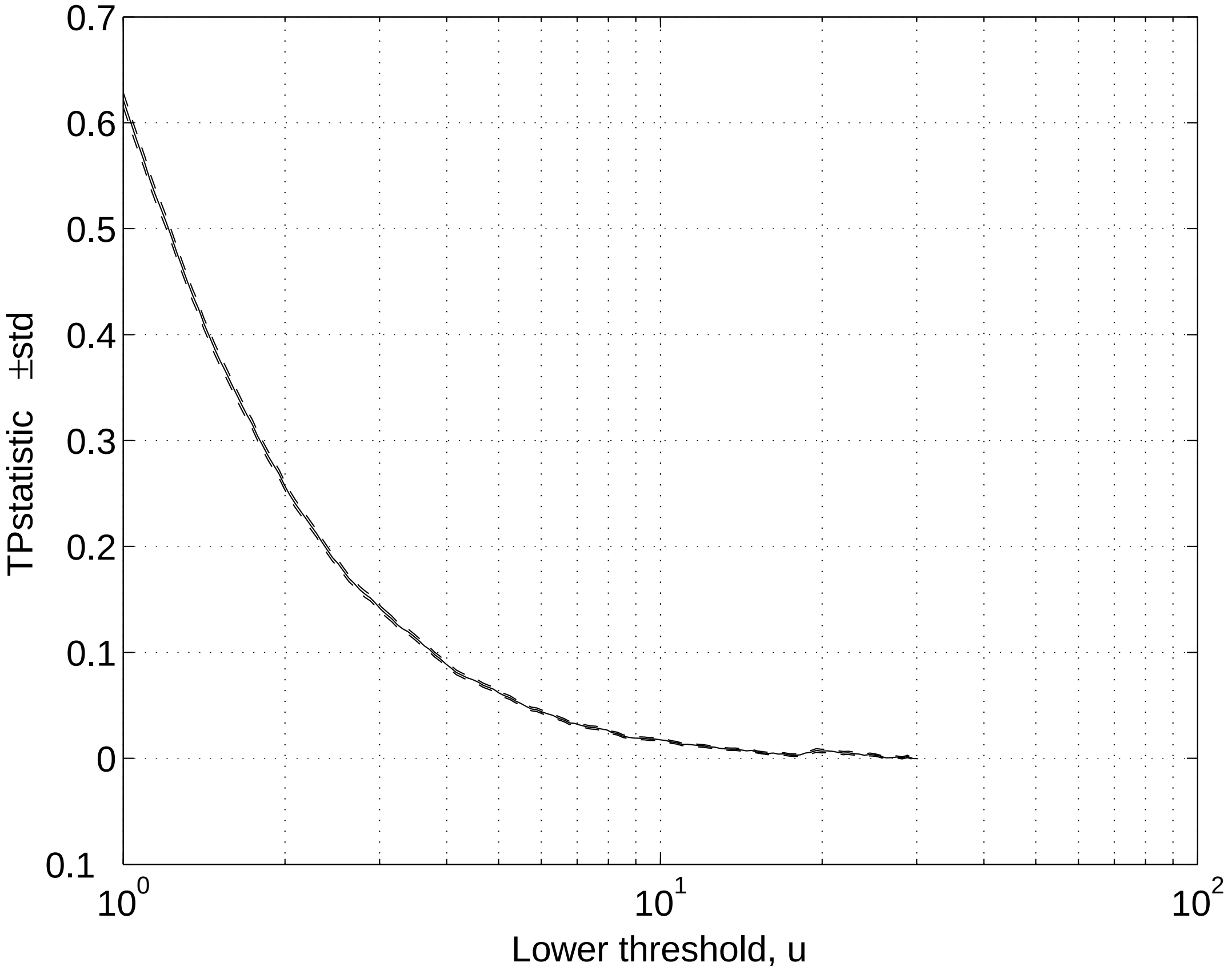

Fig. 2b

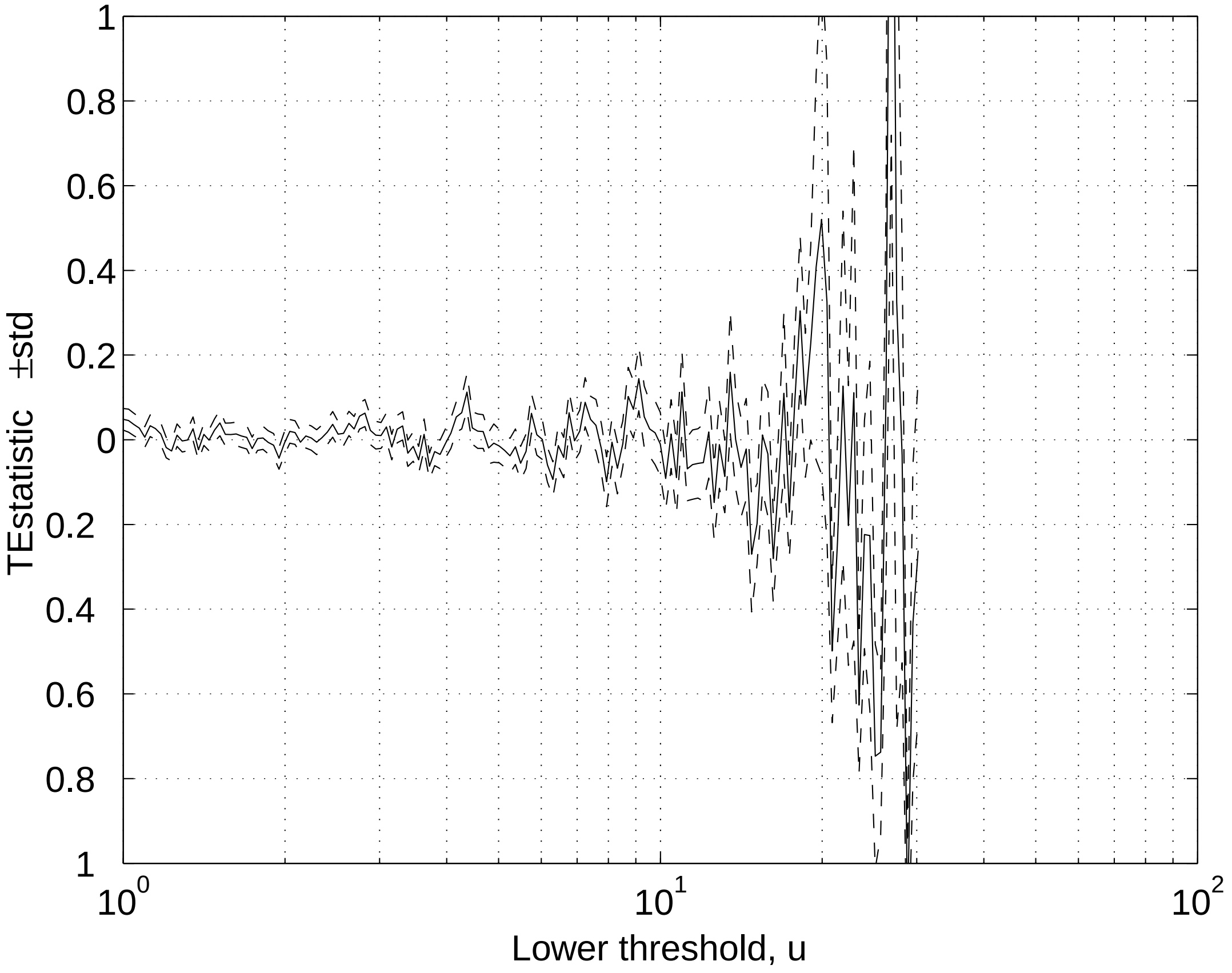

Fig. 2: a) *TP*-statistic as a function of the lower threshold *u*, applied to a simulated exponential sample of size *n=20000* with form parameter *d* = 4 generated with *u*=1 as defined in Eq.(8). The two dashed lines show plus or minus one standard deviation std estimated as exposed in the text. b) *TE*-statistic of the same synthetic data.

Fig. 3a

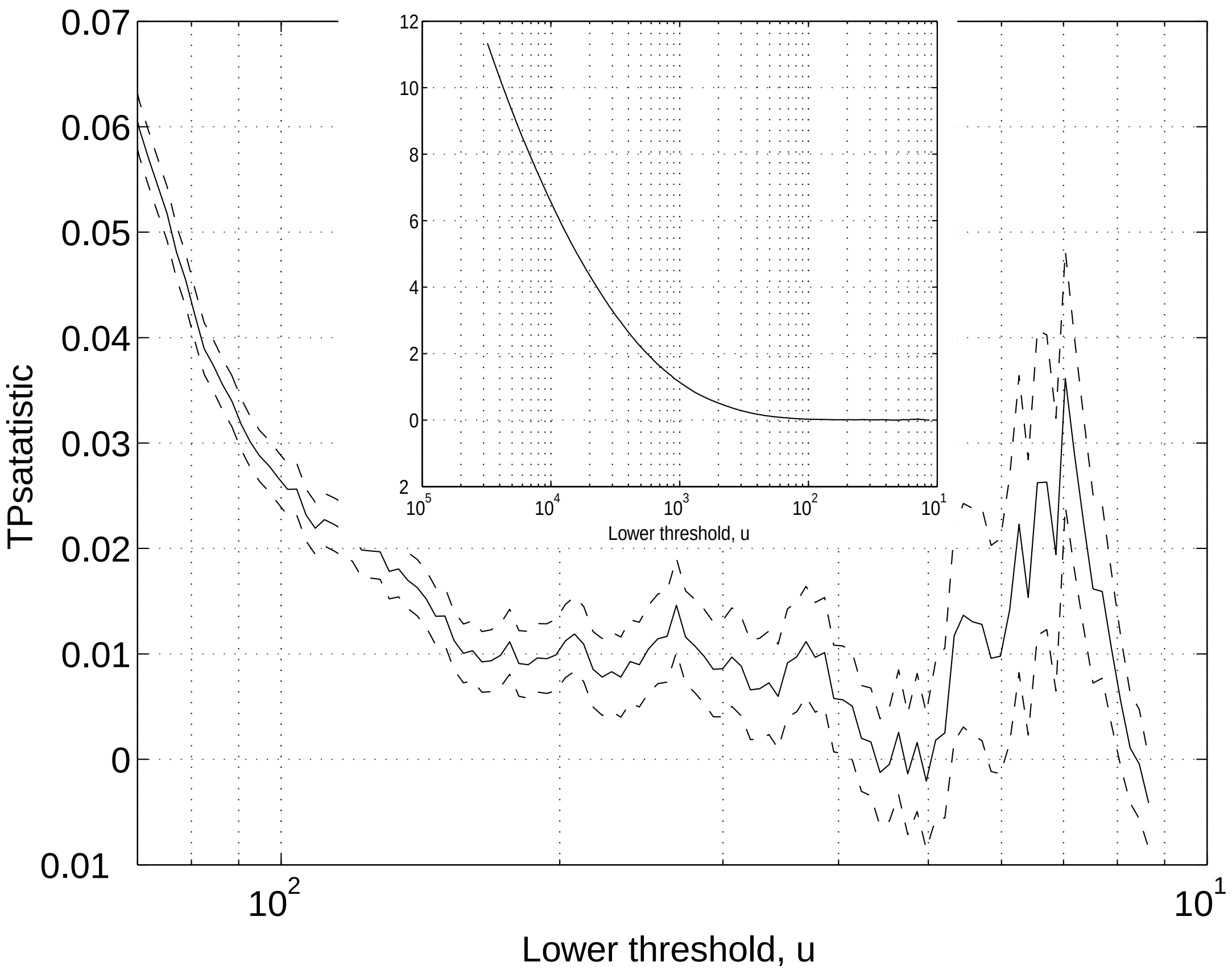

Fig. 3b

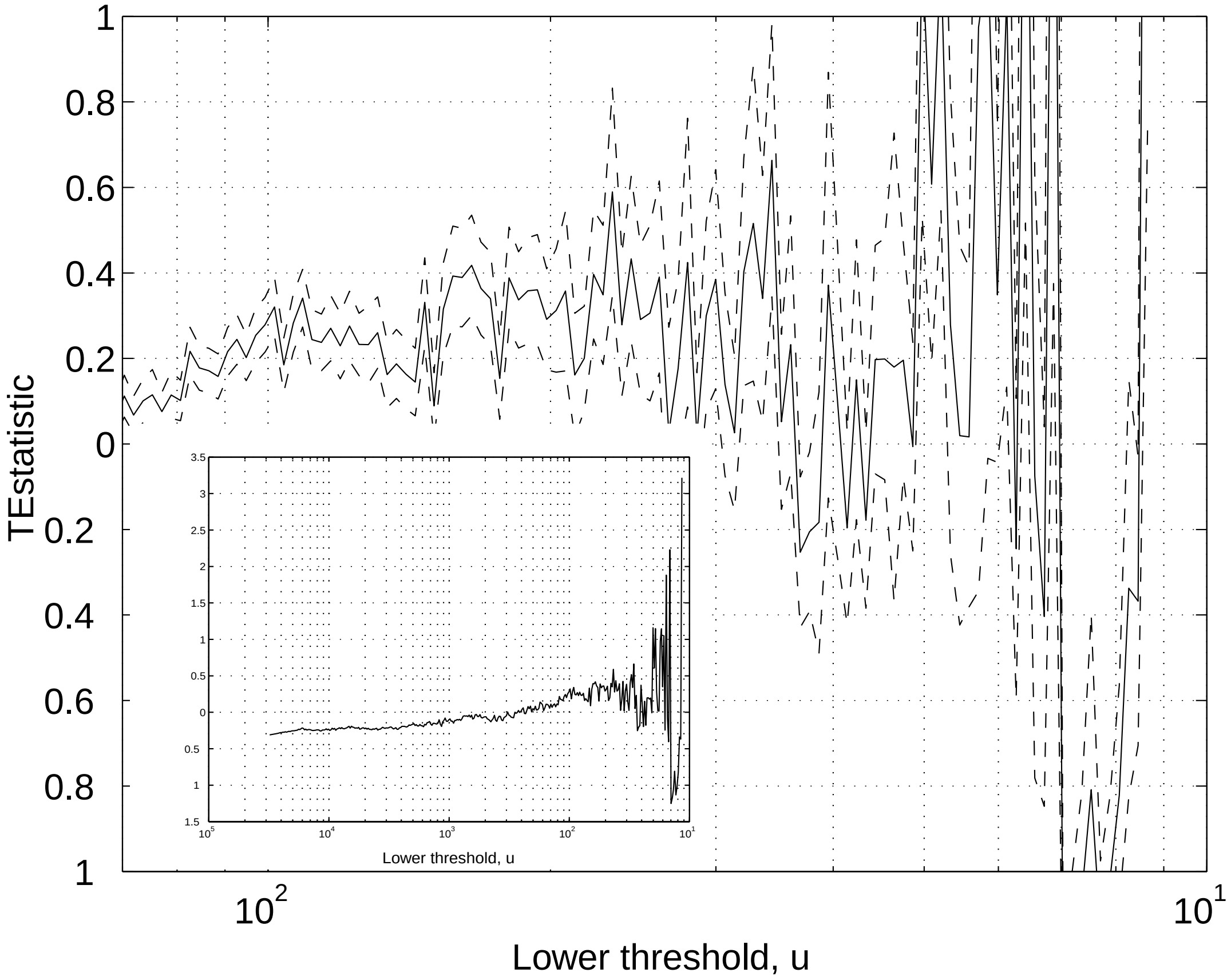

Fig. 3: *TP*-statistic *TP(u)* (a) and the *TE*-statistic *TE(u)* (b) as a function of the lower threshold *u* for positive returns of the DJIA at the daily time-scale. The two dashed lines show plus or minus one standard deviation std estimated as exposed in the text. The insets show the same statistics over the extended range of *u*.

Fig. 4a

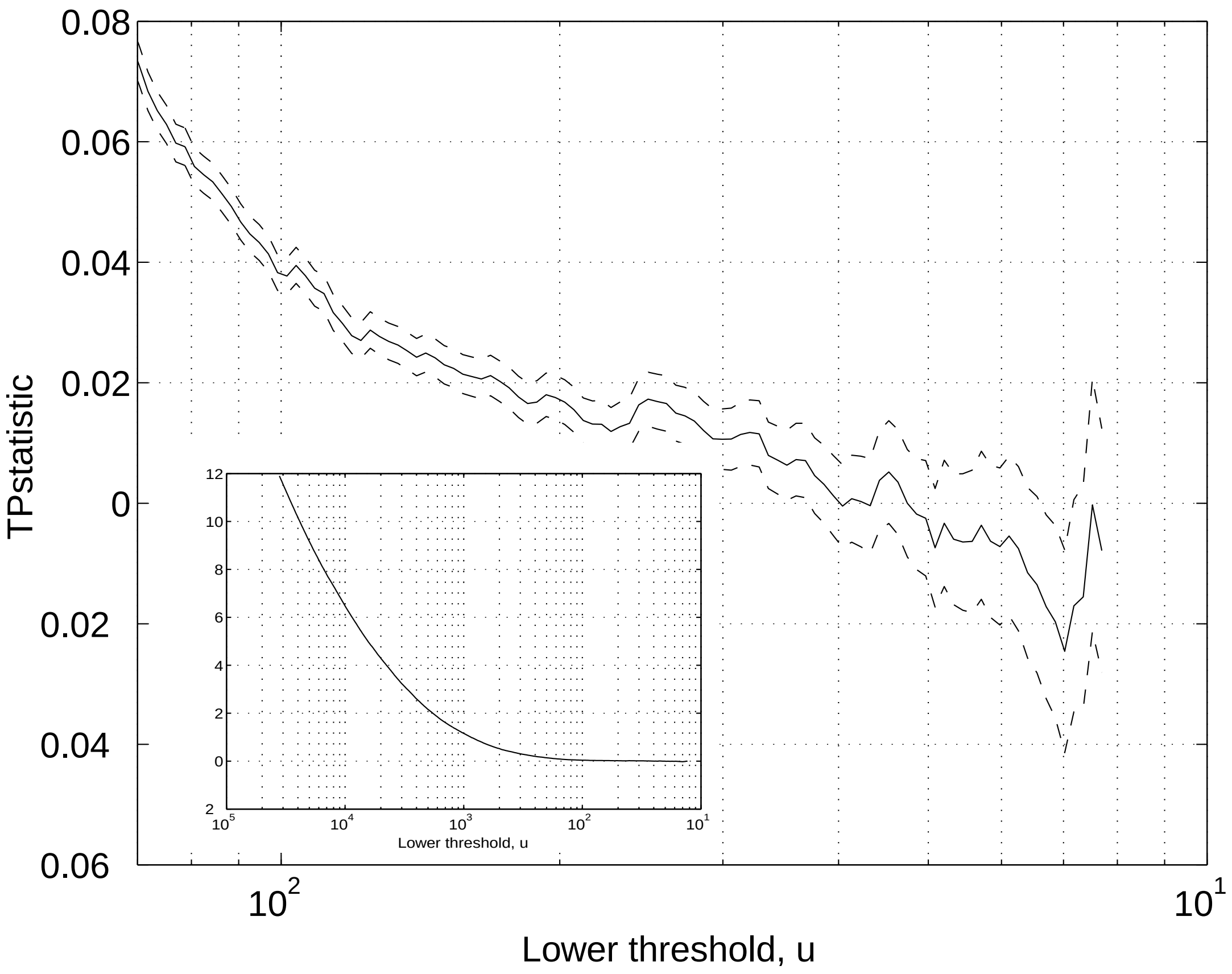

Fig. 4b

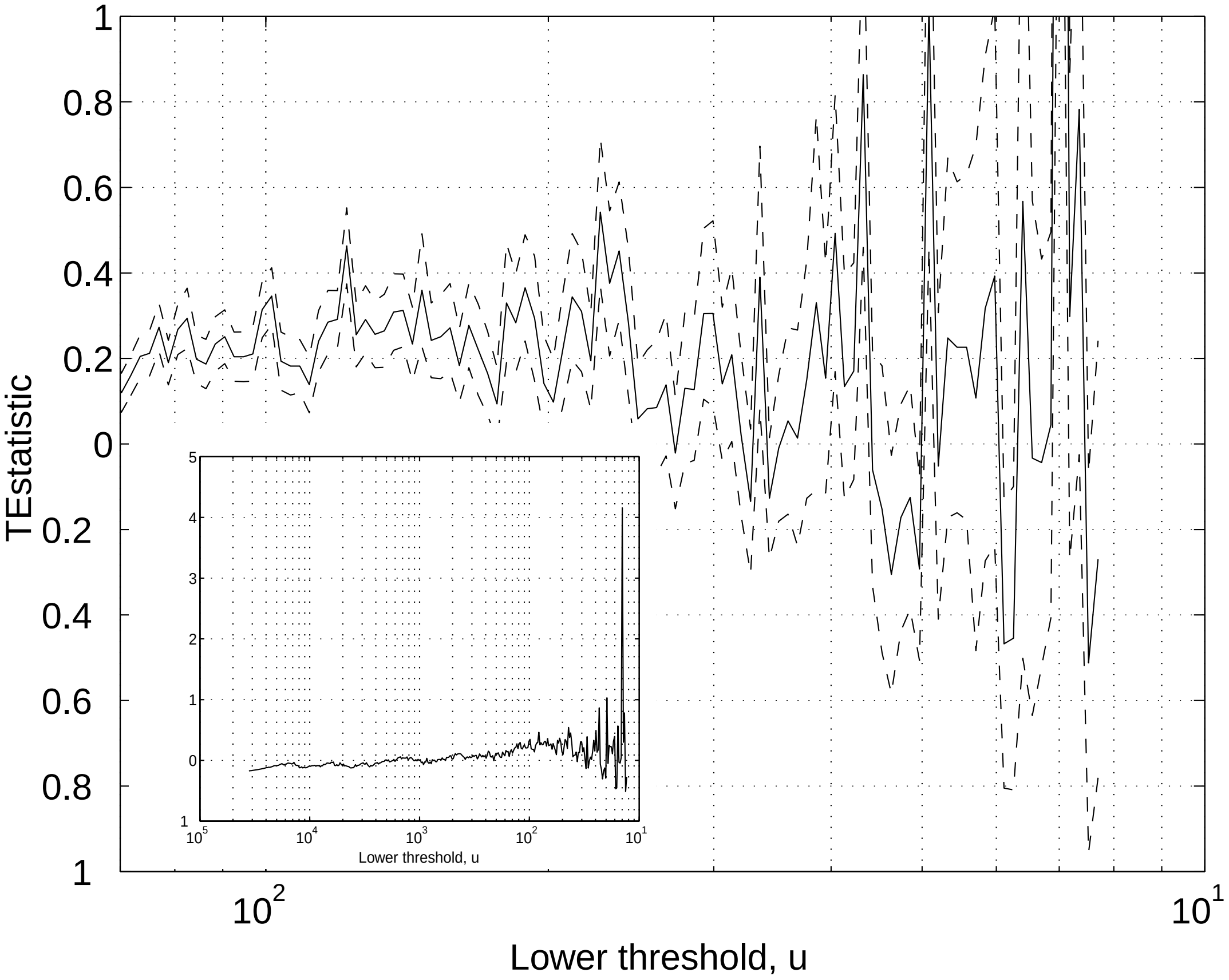

Fig. 4: *TP*-statistic *TP(u)* (a) and the *TE*-statistic *TE(u)* (b) as a function of the lower threshold *u* for negative returns of the DJIA at the daily time-scale. The two dashed lines show plus or minus one standard deviation std estimated as exposed in the text. The insets show the same statistics over the extended range of *u*.

Fig. 5a

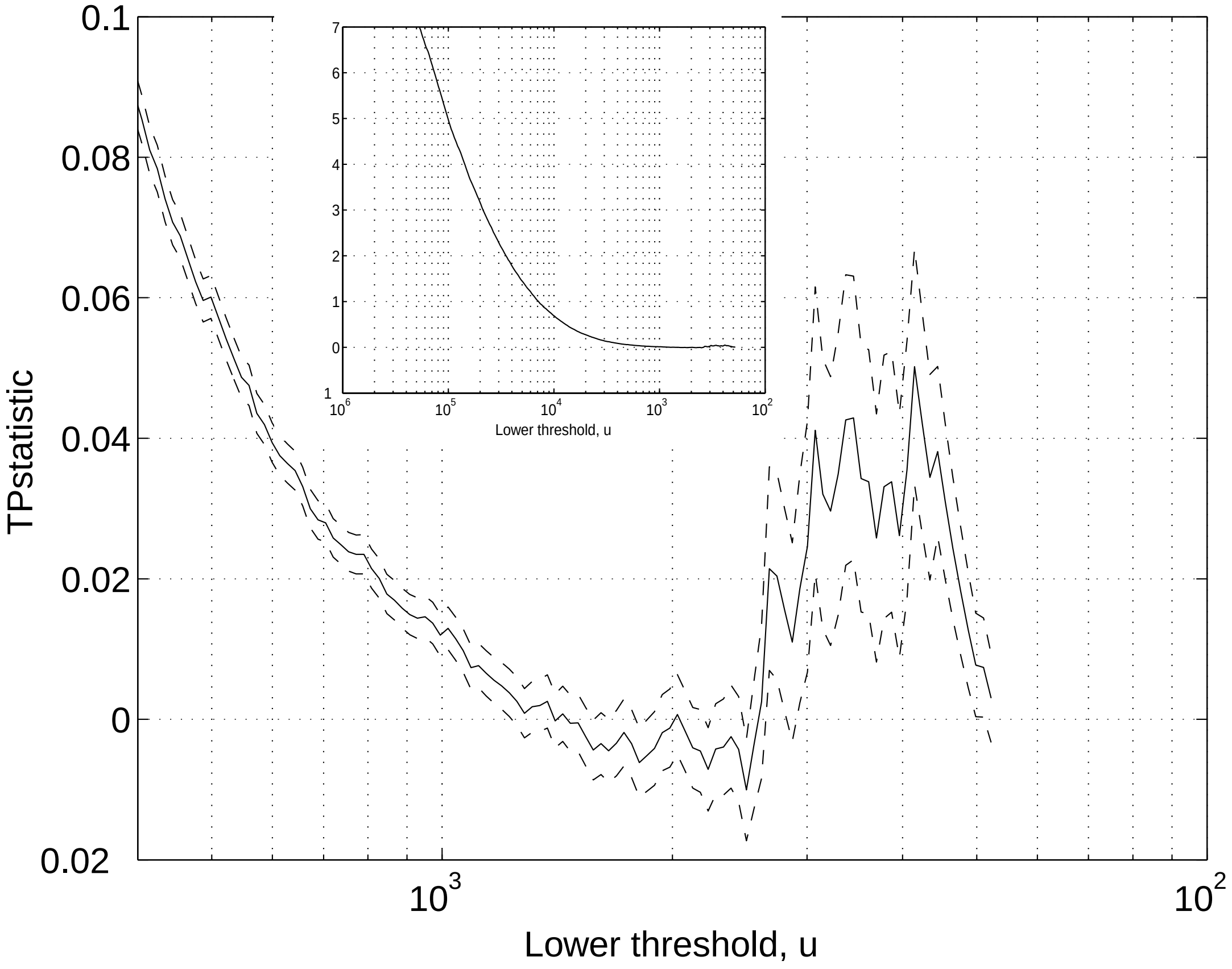

Fig. 5b

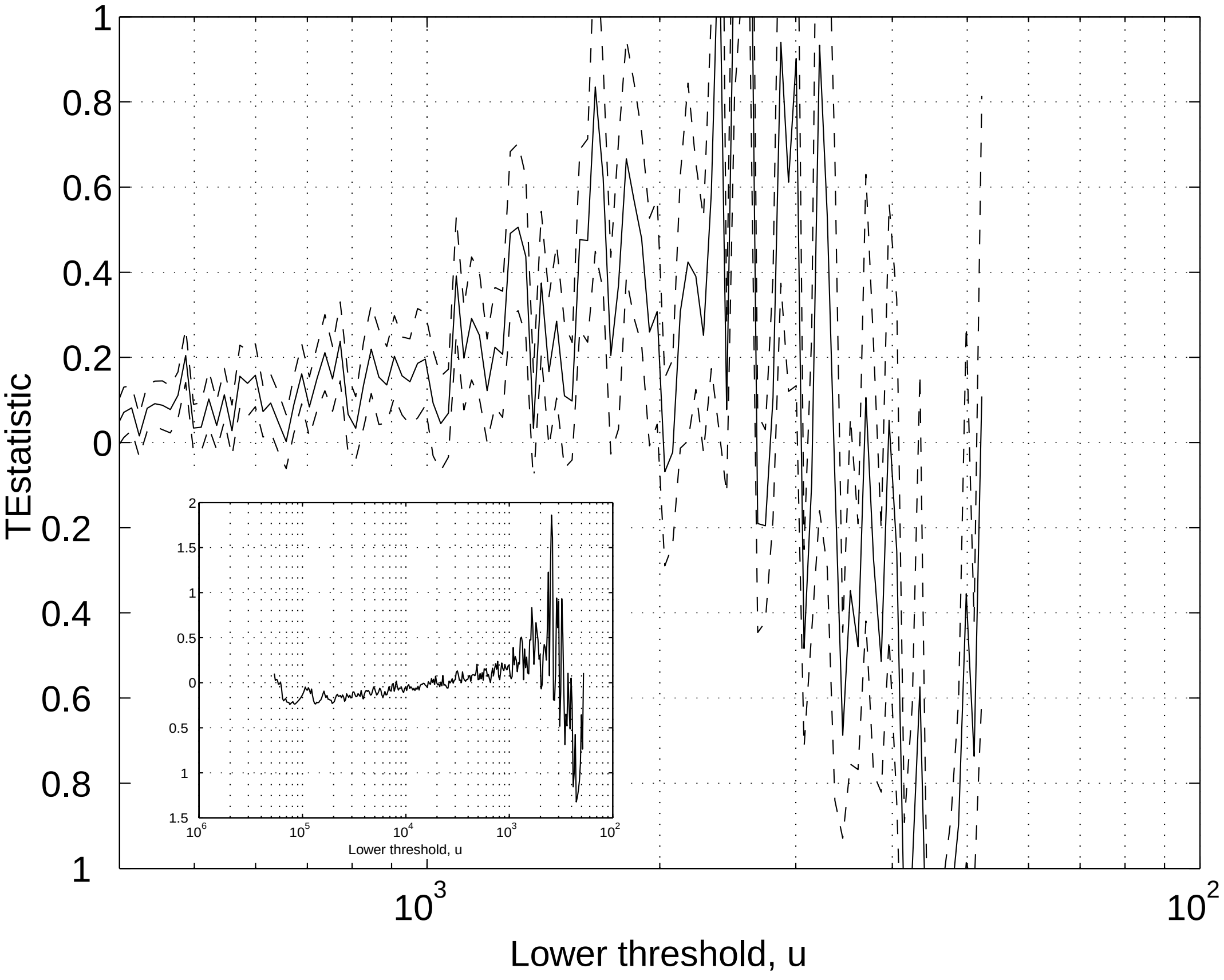

Fig. 5: *TP*-statistic *TP(u)* (a) and the *TE*-statistic *TE(u)* (b) as a function of the lower threshold *u* for positive returns of the ND at the 5-minutes time-scale. The two dashed lines show plus or minus one standard deviation std estimated as exposed in the text. The insets show the same statistics over the extended range of *u*.

Fig. 6a

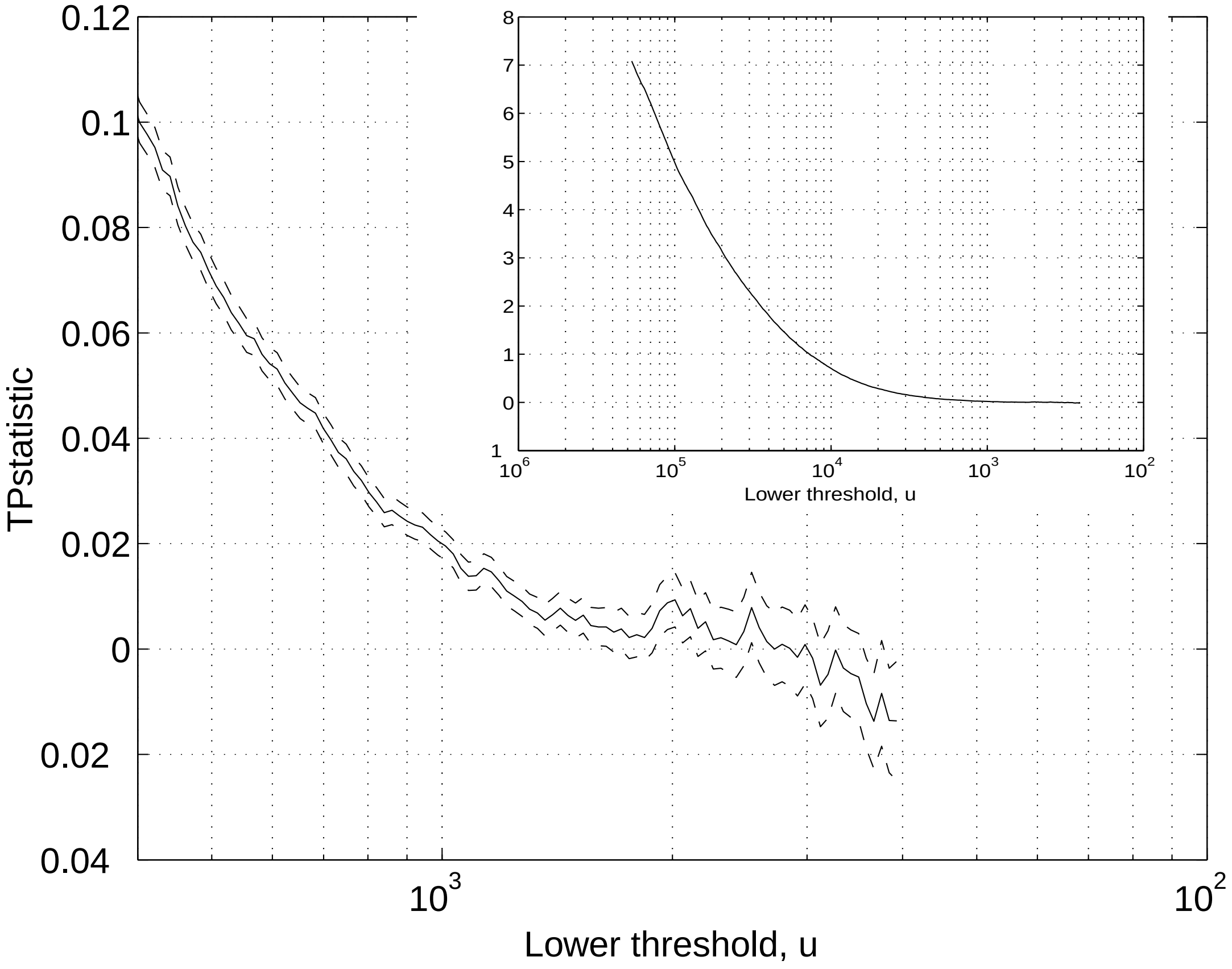

Fig. 6b

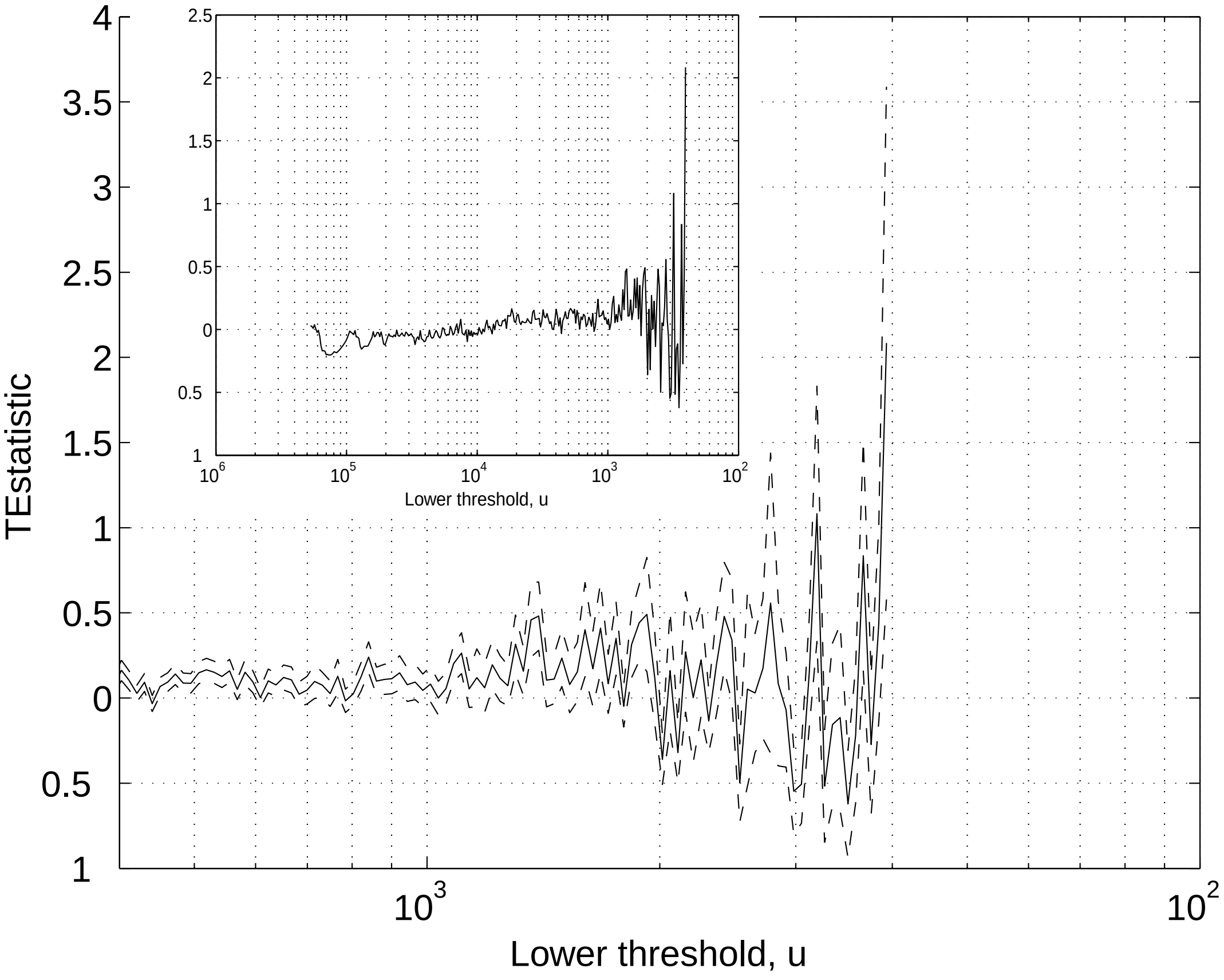

Fig. 6: *TP*-statistic *TP(u)* (a) and the *TE*-statistic *TE(u)* (b) as a function of the lower threshold *u* for negative returns of the ND at the 5-minutes time-scale. The two dashed lines show plus or minus one standard deviation std estimated as exposed in the text. The insets show the same statistics over the extended range of *u*.